\documentclass[]{JHEP3}
\usepackage{epsfig}

\def\be{\begin{equation}}
\def\ee{\end{equation}}
\def\bea{\begin{array}}
\def\eea{\end{array}}
\def\beqa{\begin{eqnarray}}
\def\eeqa{\end{eqnarray}}
\def\beqas{\begin{eqnarray*}}
\def\eeqas{\end{eqnarray*}}

\def\bp{\begin{picture}}
\def\ep{\end{picture}}
\def\bc{\begin{center}}
\def\ec{\end{center}}
\def\bfig{\begin{figure}}
\def\efig{\end{figure}}

\def\bit{\begin{itemize}}
\def\eit{\end{itemize}}
\def\nn{\nonumber}
\def\f{\frac}

\def\[{\left[}
\def\]{\right]}
\def\({\left(}
\def\){\right)}

\def\..{\left.}
\def\.{\right.}
\def\tl{\tilde}
\def\ra{\rightarrow}
\def\la{\leftarrow}

\def\tm{\times}

\def\da{\dagger}

\def\la{\lambda}

\def\al{\alpha}
\def\bt{\beta}

\def\ep{\epsilon}

\def\ga{\gamma}
\def\pa{\partial}
\def\pr{\prime}
\def\eqv{\equiv}

\title{Natural SUSY from SU(5) Orbifold GUT}

\author{Chengcheng Han$^{1,2}$, Fei Wang$^{1,2}$, Jin Min Yang$^{2}$
\\
{$^{1}$ International Joint Research Laboratory for Quantum Functional Materials of Henan Province,
and School of Physics and Engineering, Zhengzhou University, Henan 450001, P. R. China;\\
$^{2}$ State Key Laboratory of Theoretical Physics, Institute of Theoretical Physics,
       Academia Sinica, Beijing 100190, P. R. China}}

\abstract{We propose a realistic 5D orbifold GUT model that can reduce to natural (or radiative natural)
supersymmetry as the low energy effective theory. Supersymmetry as well as gauge symmetry are broken by
the twist boundary
conditions. We find that it is non-trivial to introduce other flavor symmetry other than the $SU(2)_R$
R-symmetry.  We ameliorate the tension between the small number of free parameters and
the successful electroweak symmetry breaking by introducing non-minimal Kahler potentials.
A large trilinear term $A_t$, which is necessary to give a 125 GeV Higgs boson,
is naturally provided in our scenario.  A scan under current experimental constraints shows
that our model can realize natural (or radiative natural) supersymmetry. Only radiative natural supersymmetry can naturally lead to 125 GeV higgs.
Additional dark matter species other than neutralino(like axion) are needed to provide enough relic density.
Relatively large stop masses are necessary to give realistic higg mass in most of the parameter spaces.}


\begin{document}
\maketitle
\flushbottom
\section{Introduction}
Both the ATLAS \cite{atlas} and CMS \cite{cms} collaborations have now established
the existence of a 125 GeV Higgs-like boson from the combined 7 TeV and 8 TeV LHC data.
Although the data are so far in agreement with the Standard Model (SM) prediction,
they can also be accommodated by many new physics frameworks, among which a particularly
interesting and widely studied scenario is supersymmetry (SUSY). An interesting observation
is that the mass value of the observed Higgs-like boson just falls within the narrow range
of $115-135$ GeV predicted by the Minimal Supersymmetric Standard Model (MSSM).

SUSY naturally solves the gauge hierarchy problem of the
SM and at the same time provides a viable dark matter candidate.
In SUSY, the unification of the three gauge couplings of
$SU(3)_C, SU(2)_L$ and $U(1)_Y$
at about $2\tm 10^{16}$ GeV~\cite{Ellis:1990zq} strongly suggests
the existence of Grand Unified Theories (GUTs). In addition,
the SUSY GUTs such as $SU(5)$~\cite{Georgi:1974sy} or
$SO(10)$~\cite{so10} models give us deep insights into the other SM
problems such as the emergence of the fundamental forces, the
assignment and quantization of their charges as well as the fermion masses
and mixings. Although SUSY GUTs are attractive,
it is challenging to test them at the Large Hadron Collider (LHC)
or the future International Linear Collider (ILC).

It is well known that some problems like the doublet-triplet splitting
always exist in many GUT models. One elegant way to solve this problem
is to put the GUT gauge group in the five-dimensional (5D) bulk and break the
GUT symmetry by boundary conditions, for example, by
orbifold projection. Orbifold GUT models for $SU(5)$ were proposed
in~\cite{kawamura} and then widely studied
in~\cite{at,hall,hebecker:2001wq,hebecker:2001jb,Li:2001qs, Li:2001wz,fei2,fei3,li2010}.
On the other hand, certain amounts of SUSY can be broken by assigning proper
boundary conditions to the high dimensional theory. For example, the
5D N=1 SUSY, which amounts to 4D N=2 SUSY, can be
broken to N=1 SUSY by orbifold projection.
The possibility that the remaining N=1 SUSY is broken by boundary conditions
is fairly attractive.
Such SUSY breaking mechanism is elegant and also can be interpreted to
have a dynamical origin through AdS/CFT correspondence \cite{adscft}
when such a theory is put in a Randall-Sundrum \cite{rs} type warped
extra-dimension.

We know that the SUSY partners of the SM particles can acquire masses
after SUSY breaking. Naturalness argument requires weak-scale soft SUSY parameters.
However, current collider experiments severely constrained the parameter space of the
MSSM, e.g., the LHC has pushed the first two generations of squarks above TeV-scale.
Models of natural SUSY \cite{naturalsusy} seek to retain the naturalness of SUSY
by positing a spectrum of light higgsinos and light top squarks along with very heavy
masses of other squarks and TeV-scale gluinos\footnote{In the framework of MSSM
a 125 GeV Higgs mass requires heavy stops or large $A_t$. In order to have
weak-scale stops, the MSSM must be extended, e.g., by introducing a gauge singlet
superfield (for a comparative study of low energy
SUSY models in light of the LHC Higgs data, see \cite{cao}).}.
Such models have low electroweak
fine-tuning and satisfy the LHC constraints. Besides, a relatively heavy (125 GeV) Higgs
mass indicates that natural SUSY may take the form of radiative natural SUSY \cite{rnaturalsusy}.
In this paper, we propose an SU(5) orbifold GUT model which
can reduce to the low energy natural SUSY after integrating out the heavy modes.
Like other GUT models, our scenario has only a few free parameters.

This paper is organized as follows. In Section~2
we briefly review the 4D SU(5) GUT and possible non-minimal Kahler potential extensions.
 In Section~3 we obtain the soft SUSY breaking terms from the boundary conditions.
In Section~4 we propose an approach to obtain a large trilinear coupling for stop and discuss
the relevant consistency conditions for the free parameters in our theory.
In Section~5 we scan the parameter space under current experimental constraints.
Section~\ref{sec-6} contains our conclusions.

\section{A brief review of Grand Unified Theories}
\label{sect-0}

In this Section we explain our conventions. In supersymmetric SMs,
we denote the left-handed quark doublets, right-handed up-type
quarks, right-handed down-type quarks, left-handed lepton doublets,
right-handed neutrinos and right-handed charged leptons as $Q_i$,
$U^c_i$, $D^c_i$, $L_i$, $N^c_i$, and $E^c_i$, respectively. Also,
we denote one pair of Higgs doublets as $H_u$ and $H_d$, which give
masses to the up-type quarks/neutrinos and the down-type
quarks/charged leptons, respectively.

First, we briefly review the $SU(5)$ model. We define the $U(1)_{Y}$
hypercharge generator in $SU(5)$ as follows
\begin{eqnarray}
T_{\rm U(1)_{Y}}={\rm diag} \left(-{1\over 3}, -{1\over 3}, -{1\over
3}, {1\over 2}, {1\over 2} \right)~.~\, \label{u1y}
\end{eqnarray}
Under $SU(3)_C\times SU(2)_L \times U(1)_Y$ gauge symmetry, the
$SU(5)$ representations are decomposed as follows
\begin{eqnarray}
\mathbf{5} &=& \mathbf{(3, 1, -1/3)} \oplus \mathbf{(1, 2, 1/2)}~,~ \\
\mathbf{\overline{5}} &=&
\mathbf{(\overline{3}, 1, 1/3)} \oplus \mathbf{(1, 2, -1/2)}~,~ \\
\mathbf{10} &=& \mathbf{(3, 2, 1/6)} \oplus \mathbf{({\overline{3}},
1, -2/3)}
\oplus \mathbf{(1, 1, 1)}~,~ \\
\mathbf{\overline{10}} &=& \mathbf{(\overline{3}, 2, -1/6)} \oplus
\mathbf{(3, 1, 2/3)}
\oplus \mathbf{(1, 1, -1)}~,~ \\
\mathbf{24} &=& \mathbf{(8, 1, 0)} \oplus \mathbf{(1, 3, 0)} \oplus
\mathbf{(1, 1, 0)} \oplus \mathbf{(3, 2, -5/6)} \oplus
\mathbf{(\overline{3}, 2, 5/6)}~.~\,
\end{eqnarray}
There are three families of the SM fermions whose quantum numbers
under $SU(5)$ are
\begin{eqnarray}
F'_i=\mathbf{10},~ {\overline f}'_i={\mathbf{\bar 5}},~
N^c_i={\mathbf{1}}~,~ \label{SU(5)-smfermions}
\end{eqnarray}
where $i=1, 2, 3$ for three families. The SM particle assignments in
$F'_i$ and ${\bar f}'_i$ are
\begin{eqnarray}
F'_i=(Q_i, U^c_i, E^c_i)~,~{\overline f}'_i=(D^c_i, L_i)~.~
\label{SU(5)-smparticles}
\end{eqnarray}

To break the $SU(5)$ gauge symmetry and electroweak gauge symmetry,
we introduce the adjoint Higgs field and one pair of Higgs fields
whose quantum numbers under $SU(5)$ are
\begin{eqnarray}
\Phi'~=~ {\mathbf{24}}~,~~~ h'~=~{\mathbf{5}}~,~~~{\overline
h}'~=~{\mathbf{\bar {5}}}~,~\, \label{SU(5)-1-Higgse}
\end{eqnarray}
where $h'$ and ${\overline h}'$ contain the Higgs doublets $H_u$ and
$H_d$, respectively.

Second, we briefly review the supergravity action. The bosonic piece of the supergravity action can be written as\cite{mSUGRA}
\beqa
S_B&=&\int d^4x e\[\f{M_{Pl}^2}{2}R+K_j^i(\phi^\da,\phi)(\nabla^\mu\phi_i)^\da(\nabla_\mu\phi^j)-V(\phi^\da,\phi)\.\nn\\
&+&\left.i\f{\tau}{16\pi}(F_{\mu\nu}F^{\mu\nu}+iF_{\mu\nu}\tl{F}^{\mu\nu})+h.c.\]
\eeqa
with the derivative $K_i$ of Kahler potential $K$ and Kahler metric $K_j^i$ given by
\beqa
K^i(\phi,\phi^\da)=\f{\pa K}{\pa \phi_i},~~K_j^i=\f{\pa^2 K}{\pa \phi^{j\da}\pa \phi_i}~.
\eeqa
  The supegravity scalar potential can be written
as
\beqa V(\phi,\phi^\da)=e^{\f{K}{M_{Pl}^2}}\[(K^{-1})_i^j(W^i+\f{WK^i}{M_{Pl}^2})(W_j^*+\f{W^* K_j}{M_{Pl}^2})-3\f{|W|^2}{M_{Pl}^2}\]+\f{g^2}{2}(K^iT^a\phi_i)^2 ~,~\eeqa with $W$ the superpotential.

 We consider the following non-minimal K\"ahler potential
\begin{eqnarray}
K &=&a_0 \phi_i^{\dagger} e^{2gV}\phi_i +\f{b_S}{2M_*}(S+S^\da)\phi_i^{\dagger}e^{2gV}\phi_i~.
\end{eqnarray}
  When the Higgs chiral multiplet $S$ acquires the lowest component Vaccum Expectation Value(VEV), we can have the general Kahler
potential of the form\cite{fei4}
\begin{eqnarray}
K &=&a_0 \phi_i^{\dagger}e^{2gV} \phi_i  + b_S \f{<S>}{M_*}\phi_i^{\dagger}e^{2gV}  \phi_i~,
\end{eqnarray}

 In general, SUSY-breaking scalar masses and trilinear soft terms will be obtained through F-term VEVs of $S$ in such higher dimensional
operators. Certain applications of the non-minimal Kahler potential can be found in\cite{fei5}.

\section{SUSY soft masses from Scherk-Schwarz mechanism}
We consider the five-dimensional space-time ${\cal M}_4{\tm}
S^1/Z_2$ comprising of Minkowski space ${\cal M}_4$ with
coordinates $x_{\mu}$ and the orbifold $S^1/Z_2$ with
coordinate $y \eqv x_5$. The orbifold $S^1/Z_2$ is
obtained from $S^1$ by moduling the equivalent classes:
\beqa P:~
 y{\sim} -y ~~.
\eeqa
 There are two inequivalent 3-branes
locating at $y=0$ and $y=\pi R$ which are denoted as $O$ and
$O^{\pr}$, respectively.

The five-dimensional $N=1$ supersymmetric gauge theory has 8 real supercharges,
corresponding to $N=2$ supersymmetry in four dimensions. The vector multiplet
physically contains a vector boson $A_M$ where $M=0, 1, 2, 3, 5$,
two Weyl gauginos $\lambda_{1,2}$, and a real scalar $\sigma$.
In terms of four-dimensional
$N=1$ language, it contains a vector multiplet $V(A_{\mu}, \lambda_1)$ and
a chiral multiplet $\Sigma((\sigma+iA_5)/\sqrt 2, \lambda_2)$ which
transform
in the adjoint representation of the gauge group.
The five-dimensional hypermultiplet physically has two complex scalars
$\phi$ and $\phi^c$, a Dirac fermion $\Psi$. It can be decomposed into
two 4-dimensional chiral mupltiplets $\Phi(\phi, \tl{\phi} \equiv \Psi_R)$
and $\Phi^c(\phi^c, \tl{\phi}^c \equiv \Psi_L)$, which transform as
conjugate representations of each other under the gauge group. It should be noted that $N=1$ supersymmetry in five dimensions possesses an
$SU(2)_R$ global R-symmetry under which the gauginos from the vector
multiplets $(\la_1,\la_2)$ and complex scalars $(\phi,\phi^{c\da})$ from hypermultiplets form $SU(2)_R$ doublets.

The general action\cite{nima2} for the gauge fields and their
couplings to the bulk hypermultiplet $\Phi$ is given by
\begin{eqnarray}
S&=&\int{d^5x}\frac{1}{k g^2}
{\rm Tr}\left[\frac{1}{4}\int{d^2\theta} \left(W^\alpha W_\alpha+{\rm H.
C.}\right)
\right.\nonumber\\&&\left.
+\int{d^4\theta}\left((\sqrt{2}\partial_5+ {\bar \Sigma })
e^{-V}(-\sqrt{2}\partial_5+\Sigma )e^V+
\partial_5 e^{-V}\partial_5 e^V\right)\right]
\nonumber\\&&
+\int{d^5x} \left[ \int{d^4\theta} \left( {\Phi}^c e^V {\bar \Phi}^c +
{\bar \Phi} e^{-V} \Phi \right)
\right.\nonumber\\&&\left.
+ \int{d^2\theta} \left( {\Phi}^c (\partial_5 -{1\over {\sqrt 2}} \Sigma)
\Phi + {\rm H. C.}
\right)\right]~.~\,
\label{VD-Lagrangian}
\end{eqnarray}
   The gauge symmetry and supersymmetry can be broken by choosing suitable
representations for $P$.
For a  field $\phi$ in
the representation of unbroken gauge symmetry, we obtain
its transformation rules as follows
\beqa
\phi(x_{\mu},y)& \to &  \phi(x_{\mu},-y) =  \eta_{\phi} \phi(x_{\mu},y)
~,~\,
\eeqa
where $\eta_{\phi}=\pm1$.

Because the action is invariant under the parity $P$,
 we obtain the transformation rules of the
vector multiplet under the parity operator $P$
\begin{eqnarray}
V(x^{\mu},y)&\to  V(x^{\mu},-y) = P V(x^{\mu}, y) P^{-1}
~,~\,
\end{eqnarray}
\begin{eqnarray}
 \Sigma(x^{\mu},y) &\to\Sigma(x^{\mu},-y) = - P \Sigma(x^{\mu},
y) P^{-1}~.~\,
\end{eqnarray}

If the hypermultiplet belongs to the fundamental or anti-fundamental
representations, note that $P=P^{-1}$, we have
\begin{eqnarray}
\Phi(x^{\mu},y)&\to \Phi(x^{\mu}, -y)  = \eta_{\Phi} P
\Phi(x^{\mu},y)
~,~\,
\end{eqnarray}
\begin{eqnarray}
\Phi^c(x^{\mu},y) &\to \Phi^c(x^{\mu}, -y)  = -\eta_{\Phi} P
\Phi^c(x^{\mu},y) ~.~\,
\end{eqnarray}
where $\eta_{\Phi} = \pm 1$.
And if the hypermultiplet belongs to the symmetric, anti-symmetric
or adjoint representations,  we have
\begin{eqnarray}
\Phi(x^{\mu},y)&\to \Phi(x^{\mu}, -y)  = \eta_{\Phi} P
\Phi(x^{\mu},y) P
~,~\,
\end{eqnarray}
\begin{eqnarray}
\Phi^c(x^{\mu},y) &\to \Phi^c(x^{\mu}, -y)  = -\eta_{\Phi} P
\Phi^c(x^{\mu},y) P ~.~\,
\end{eqnarray}
where $\eta_{\Phi} = \pm 1$.
 We can also introduce non-trivial twisting boundary conditions. The non-trivial twist $T$ for translation
 \beqa
T \phi(x_\mu,y)=\phi(x_\mu,y+2\pi R)~,~~~~~
 \eeqa
should satisfy the following consistency condition\cite{quiros,nomura} between the orbifolding $P^{\pr\pr}$ and the translation
\beqa
\label{bconsistent}
 TP^{\pr\pr}T=P^{\pr\pr}~.~~~~
\eeqa

We can simultaneously impose the gauge symmetry breaking as well as supersymmetry breaking boundary conditions in our scenario.
We impose trivial orbifold projection conditions however non-trivial twisting to break the gauge symmetry.
The nontrivial twisting
\beqa
V(x_\mu,y+2\pi R)&=& T V(x_\mu,y) T^{-1},\\
H(x_\mu,y+2\pi R)&=& T H(x_\mu,y).
\eeqa
can break the gauge symmetry in certain fixed points of the orbifold.
An alternative way to see the gauge symmetry breaking by boundary conditions is that the translation twist $T$ and reflection $P_{0}$ at $y=0$
can be combined to give the reflection at $y=\pi R$ with the reflection operator
 \beqa
 P_{\pi R}{\equiv} TP_{0}: ~y+ \pi R \ra -y+ \pi R.
 \eeqa
 Then the massless zero modes can preserve different gauge symmetries by assigning proper $(P_0, P_{\pi R})$ boundary conditions to the two fix points.
Only $\phi_{++}(x_{\mu},y)$ possess a four-dimensional massless zero mode.
It is easy to see that $\phi_{++}$ and $\phi_{+-}$ are non-vanishing
at $y=0$ brane and $\phi_{++},\phi_{-+}$ are non-vanishing at $y=\pi R$ brane.

We choose
\beqa
P_0=(+1,+1,+1,+1,+1),~~T=(+1,+1,+1,-1,-1),
\eeqa
so that the boundary conditions preserves SU(5) gauge symmetry in the $y=0$ brane as well as in the bulk while breaks the SU(5) gauge symmetry down to $SU(3)_c\times SU(2)_L\times U(1)_Y$
in the $y={\pi R}$ brane. In our scenario, the undesirable doublet-triplet splitting problem is solved by orbifold projection.
Besides, the boundary conditions will break $N=2$ supersymmetry to $N=1$ supersymmetry.

The remaining $N=1$ supersymmetry breaking can be realized via the Scherk-Schwarz mechanism
through the boundary conditions \cite{scherkschwarz,susyboundary}.  In fact, Scherk-Schwarz mechanism can be understood in a dynamical way from the VEVs of the auxiliary component field of the radion superfield\cite{pomarol}. That is, supersymmetry from Scherk-Schwarz mechanism can be interpreted as spontaneous breaking of local supersymmetry through a Wilson line in the supergravity completion of the theory\cite{gersdorff}. The radion field can be embedded into an radion superfield
\beqa
T=R+iB_5+\theta \Psi_R^5+\theta^2 F_T,
\eeqa
with $B_5$ the fifth component of the graviphoton and $\Psi_R^5$ the fifth component of the right-handed gravitino.
The nonzero F-term $F_T$ of radion superfield will be responsible for supersymmetry breaking. From the action with radion in flat extra dimension
\beqa
\Delta S_5\supset \int d^5x \[\f{1}{4g_5^2}\int d^2\theta T W^a W^a+h.c.\]+\cdots,
\eeqa
and taking $<T>=R+\theta^2 F_T$, we will obtain\cite{pomarol} the 5D Majorana gaugino masses
\beqa
{\cal L}\supset \f{1}{2}\f{F_T}{2R}\la_i^T C\la_i,
\eeqa
and subsequently the mass terms for zero modes. Soft mass terms for sfermions will be similarly obtained. This is the same spectrum as the one obtained from Scherk-Schwarz mechanism with an R-symmetry. The VEVs of $T$ (and $F_T$) can be determined by the stabilization mechanism for the radion, for example, the Goldberger-Wise like mechanism\cite{GW-mechanism}. Non-vanishing  $F_T$ term VEVs will in general lead to non-vanishing F-terms for hypermultiplet field.
With possible higher-dimensional operators in our scenario, F-term VEVs($F_\phi$ and possibly $F_T$) will lead to new contributions to soft supersymmetry breaking parameters. However, the contributions from singlet $S$ to soft supersymmetry breaking parameters are always accompanied by an additional suppression factor $<S>/M_*$ and becomes subleading. Therefore, we will not consider such contributions in our scenario. In fact, our scenario may be best understood to assume certain radion stabilization mechanisms which yield non-vanishing F-component of radion and at the same time those negligibly vanishing for the other singlets.

 The boundary conditions for vector multiplets $[V(A_{\mu}, \lambda_1),\Sigma((\sigma+iA_5)/\sqrt 2, \lambda_2)]$ and hypermultiplets $[\Phi(\phi, \tl{\phi}), \Phi^c(\phi^c, \tl{\phi}^c)]$ can be written as
\beqa
\(\bea{c}
V\\
\Sigma
\eea\)(x_\mu,-y)&=&\(\bea{c}
V\\
-\Sigma
\eea\)(x_\mu,y),\nn\\
\(\bea{cc}
\Phi_1&\Phi_2\\
\Phi_1^c&\Phi_2^c
\eea\)(x_\mu,-y)
&=&\(\bea{cc}
\Phi_1&\Phi_2\\
\Phi_1^c&\Phi_2^c
\eea\)(x_\mu,y).
\eeqa
for reflection with respect to $y=0$ and
 \beqa
A^M(x^\mu,y+2\pi R)&=& T A^M(x^\mu,y) T ~,~ \\
\sigma(x^\mu,y+2\pi R)&=& T \sigma(x^\mu,y) T ~,~ \\
\left(\bea{c} \la_1  \\
\la_2\eea \right)(x^\mu,y+2\pi R)&=&e^{-2\pi i\al \sigma_2} T\left(\bea{c} \la_1\\
\la_2\eea \right)(x^\mu,y) ~,~ ~\,\\
 \left(\bea{cc} \phi_1~&\phi_2  \\
\phi_1^{c\da}&\phi_2^{c\da}\eea \right)(x^\mu,y+2\pi R)&=&e^{-2\pi i\al \sigma_2}T\left(\bea{cc} \phi_1~&\phi_2  \\
\phi_1^{c\da}&\phi_2^{c\da}\eea \right)(x^\mu,y)~,\,\\
\left(\bea{cc} \tl{\phi}_1~&\tl{\phi}_2  \\
\tl{\phi}_1^{c\da}&\tl{\phi}_2^{c\da}\eea \right)(x^\mu,y+2\pi R)&=&T\left(\bea{cc}  \tl{\phi}_1~&\tl{\phi}_2  \\
\tl{\phi}_1^{c\da}&\tl{\phi}_2^{c\da}\eea \right)(x^\mu,y)~,\,
\eeqa
 for twisting.
The consistency conditions for twisting
\beqa
\label{bconsistent}
 TP^{\pr\pr}T=P^{\pr\pr}~,~~~~
\eeqa
requires
 \beqa T=\exp\(-2\pi i \sigma_2\alpha\)~,~~~ \eeqa
for non-trivial orbifolding projection
\beqa P^{\pr\pr} =\sigma_3~,~~~~~~~~ \eeqa
of $SU(2)_R$ global symmetry. For trivial boundary conditions of reflection at $y=0$ with respect to gauge symmetry,
the consistency conditions only require $T^2=1$ which is trivially satisfied in our scenario.
Squarks and sleptons of the first two generations can have non-trivial
boundary conditions. In order to give low energy matter spectrum of MSSM, we have to introduce
two hypermultiplets $10^A,10^B$( $\bar{5}^A,\bar{5}^B$) for each type of representations in SU(5).
The decomposition of the matter content with respect to the orbifolding in terms of
the $(P_0,P_{\pi R})$ parity assignment is
\beqa
{\bf 10}^A_i&=&{\bf (\bar{3},1)_{-4/3}^{(+,+)}\oplus(3,2)_{1/3}^{(+,-)}\oplus(1,1)_{2}^{(+,+)}},\\
{\bf 10}^B_i&=&{\bf (\bar{3},1)_{-4/3}^{(+,-)}\oplus(3,2)_{1/3}^{(+,+)}\oplus(1,1)_{2}^{(+,-)}},\\
{\bf \bar{5}^A}_i&=&{\bf (\bar{3},1)_{2/3}^{(+,+)}\oplus(1,2)_{-1}^{(+,-)}},\\
{\bf \bar{5}^B}_i&=&{\bf (\bar{3},1)_{2/3}^{(+,-)}\oplus(1,2)_{-1}^{(+,+)}},
\eeqa
with $i=1,2$ being the family index. There is a $SU(2)_V$ global symmetry among these hypermultiplets with both $(10^A,10^B)$ and $(\bar{5}^A,\bar{5}^B)$ transform as $SU(2)_V$ doublets
\beqa
\(\bea{cc}10^A_i& 10^B_i\\\overline{10}^A_i& \overline{10}^B_i\eea\)(x^\mu,y+2\pi R)&=& \(\bea{cc}10^A_i& 10^B_i\\\overline{10}^A_i& \overline{10}^B_i\eea\)(x^\mu,y)e^{2\pi i\gamma \sigma_2},\nn\\
\(\bea{cc}\bar{5}^A_i& \bar{5}^B_i\\{5}^A_i& 5^B_i\eea\)(x^\mu,y+2\pi R)&=& \(\bea{cc}\bar{5}^A_i& \bar{5}^B_i\\{5}^A_i& 5^B_i\eea\)(x^\mu,y)e^{2\pi i\gamma \sigma_2}.
\eeqa
where the subscript $i=1,2$ denotes generation index.
In our scenario, we place the first two generation matter contents on the bulk while the
third generation matter contents on the $y=0$ brane. This scenario will lead to natural SUSY
in the IR limit after we integrate out the heavy modes.

In SUSY SU(5) GUT, we need two hypermultiplets $H_1$ and $H_2$ (denoted by $\Phi_1$ and $\Phi_2$, respectively) in
${\bf 5}$ representation of SU(5) GUT group to give low energy two higgs doublets
$h_u,h_d$ after orbifolding. We find that it is non-trivial to adopt other global symmetries in additional to
the $R$-symmetry $SU(2)_R$ in our scenario because of the Yukawa couplings.
In order to accommodate both the global symmetry for the Higgs sector and the Yukawa couplings
between the first two generation matter contents and the Higgs field, we must introduce additional
Higgs hypermultiplet field $\Phi_3$ to ensure the Higgs sector transforms in ${\bf 3}$ representation
of the $SU(2)_V$ symmetry. The ${\bf 3}$ representation of $SU(2)_V$ can be generated by an $SO(3)$ generator
\beqa
T^\pr=\exp\(2\pi i\sum\limits_{a=1}^3T^a\theta^a\),
\eeqa
with
\beqa
T^1=\(\bea{ccc}&-i&\\i&&\\&&0\eea\),~~T^2=\(\bea{ccc}&&i\\&0&\\-i&&\eea\),~~T^3=\(\bea{ccc}0&&\\&&-i\\&i&\eea\).
\eeqa
The consistency condition $T^\pr P^\pr T^\pr=P^\pr$ requires
\beqa
\{P^\pr,\sum\limits_{a}T_a\theta^a\}=0.
\eeqa
We require $\phi_i$ adopt the following non-trivial boundary conditions
\beqa
P_0(\Phi_1)=(+1,+1,+1,+1,+1)~,~~~~P_{\pi R}(\Phi_1)=(-1,-1,-1,+1,+1),\nn\\
P_0(\Phi_2)=(+1,+1,+1,+1,+1)~,~~~~P_{\pi R}(\Phi_2)=(+1,+1,+1,-1,-1),\nn\\
P_0(\Phi_3)=(-1,-1,-1,-1,-1)~,~~~~P_{\pi R}(\Phi_3)=(+1,+1,+1,-1,-1).
\eeqa
Then the most general flavor rotation that can be compatible with the projection
\beqa
P^\pr_0=(+1,+1,-1)~,~~P^\pr_{\pi R}=(-1,+1,+1),
\eeqa
are given by
\beqa
T^\pr=\exp\(2\pi i[T^2\theta^2+T^3\theta^3]\).
\eeqa

The boundary conditions for the Higgs sector can be written as
\beqa
\left(\bea{ccc} \phi_1~&\phi_2~&\phi_3  \\
\phi_1^{c\da}&\phi_2^{c\da}&\phi_3^{c\da}\eea \right)(x^\mu,y+2\pi R)&=&e^{-2\pi i\al \sigma_2}T\left(\bea{ccc} \phi_1~&\phi_2~&\phi_3 \nn \\
\phi_1^{c\da}&\phi_2^{c\da}&\phi_3^{c\da}\eea \right)(x^\mu,y)e^{2\pi i[\theta^2 T^2+\theta^3 T^3]} ~,\,\\
\left(\bea{ccc} \tl{\phi}_1~&\tl{\phi}_2~&\tl{\phi}_3  \\
\tl{\phi}_1^{c\da}&\tl{\phi}_2^{c\da}&\tl{\phi}_3^{c\da}\eea \right)(x^\mu,y+2\pi R)&=&T\left(\bea{ccc}  \tl{\phi}_1~&\tl{\phi}_2 ~&\tl{\phi}_3 \\
\tl{\phi}_1^{c\da}&\tl{\phi}_2^{c\da}&\tl{\phi}_3^{c\da}\eea \right)(x^\mu,y)e^{2\pi i[\theta^2 T^2+\theta^3 T^3]}.
\eeqa
The relation between the parameter $\theta$ and $\gamma$ is determined by the relation
\beqa
U\sigma^a U^{-1}=\sum\limits_{a}\sigma^b R_{ab}.
\eeqa
Using the expression
\beqa
U=e^{2\pi i \gamma \sigma_2}
\eeqa
and the commutation relation for $\sigma_a$, we can obtain the relation
between $\gamma$ and $\theta^i$
\beqa
\theta^2=2\gamma~,~~~~~~~~~~\theta^3=0.
\eeqa
So the transformation law for the Higgs fields, which is compatible with the global symmetry,
can be written as
\beqa
R=e^{4\pi i \gamma T^2}.
\eeqa
It is well know that CP violation constraints require heavy superpartners for light quarks. In order to give heavy sparticle masses to the first two generation matter contents, We introduce a global $SU(2)_F$ family symmetry for the first two generations with the third family being $SU(2)_F$ singlet. Complete model with $SU(2)_F$ family symmetry were discussed in ref.\cite{su2-global}. (Discrete) version of $SU(2)_H$ model involving additional gauged $U(1)$ to reproduce natural supersymmetry are discussed in ref.\cite{pomarol-ns}.
The boundary conditions in our scenario can be written as
\beqa
\(\bea{cc}10^{A,B}_1& 10^{A,B}_2\\\overline{10}^{A,B}_1& \overline{10}^{A,B}_2\eea\)(x^\mu,y+2\pi R)&=& \(\bea{cc}10^{A,B}_1& 10^{A,B}_2\\\overline{10}^{A,B}_1& \overline{10}^{A,B}_2\eea\)(x^\mu,y)e^{2\pi i\beta \sigma_2},\nn\\
\(\bea{cc}\bar{5}^{A,B}_1& \bar{5}^{A,B}_2\\{5}^{A,B}_1& 5^{A,B}_2\eea\)(x^\mu,y+2\pi R)&=& \(\bea{cc}\bar{5}^{A,B}_1& \bar{5}^{A,B}_2\\{5}^{A,B}_1& 5^{A,B}_2\eea\)(x^\mu,y)e^{2\pi i\beta \sigma_2},
\eeqa
with $\bt$ the parameter for $SU(2)_{F}$ flavor symmetry. The higgs fields are singlet with respect to this family symmetry, so they receive no contributions from this twisting.
General renormalizable yukawa couplings compatible with $SU(2)_F$ that contained within the $SU(2)_V$ invariant Lagrangian can be introduced as
 \beqa
W&\supseteq& y^{u}\epsilon_{ij}{\bf 10}^{A,i}{\bf 10}^{B,j}{\bf 5}_H+y^d\epsilon_{ij}{\bf 10}^{B,i}{\bf \bar{5}}^{A,j}{\bf \bar{5}}_H+y^e\epsilon_{ij}{\bf 10}^{A,i}{\bf \bar{5}}^{B,j}{\bf \bar{5}}_H\nn\\
&+&y^{u}_{33}{\bf 10}^{A,s}{\bf 10}^{B,s}{\bf 5}_H+y^d_{33}{\bf 10}^{B,s}{\bf \bar{5}}^{A,s}{\bf \bar{5}}_H+y^e_{33}{\bf 10}^{A,s}{\bf \bar{5}}^{B,s}{\bf \bar{5}}_H.
\eeqa
 In this formula, the indices $'i,j'$ denote the first two generations while the index $'s'$ denotes the third generation. It is obvious that there are no mixing between the third generation
and the first two in the previous Lagrangian. Thus to reproduce the Cabbibo-Kabayashi-Maskawa(CKM) mixing, we need to introduce non-renormalizable operators compatible with the $SU(2)_F$ family symmetry.
 We can introduce two $SU(2)_F$ doublet higgs fields $\Phi^{\al,i}(\al=1,2)$ with $'i'$ the $SU(2)_F$ index to generate the desired high-dimensional operators
\beqa
W &\supseteq & \f{1}{M_{Pl}}\[y^u_\al\epsilon_{ij} {\bf 10}^{A,i}{\bf 10}^{B,s}\Phi^{\al,j}{\bf 5}_H+y^{u\pr}_\al\epsilon_{ij} {\bf 10}^{A,s}{\bf 10}^{B,i}\Phi^{\al,j}{\bf 5}_H+y^{d}_\al\epsilon_{ij}{\bf 10}^{B,i}{\bf \bar{5}}^{A,s}\Phi^{\al,j}{\bf \bar{5}}_H\.\nn\\
&&~~~~+\left.\f{}{} y^{d\pr}_\al\epsilon_{ij}{\bf 10}^{B,s}{\bf \bar{5}}^{A,i}\Phi^{\al,j}{\bf \bar{5}}_H
+y^{e}_\al\epsilon_{ij}{\bf 10}^{A,i}\Phi^{\al,j}{\bf \bar{5}}^{B,s}{\bf \bar{5}}_H+y^{e\pr}_\al\epsilon_{ij}{\bf 10}^{A,s}{\bf \bar{5}}^{B,i}\Phi^{\al,j}{\bf \bar{5}}_H\]\nn\\
&+&\f{1}{M_{Pl}^2}\[y^u_{\al\bt}\epsilon_{il}\epsilon_{jm}{\bf 10}^{A,i}{\bf 10}^{B,j}\Phi^{\al,l}\Phi^{\bt,m}{\bf 5}_H+
y^d_{\al\bt}\epsilon_{il}\epsilon_{jm}{\bf 10}^{A,i}{\bf \bar{5}}^{A,j}\Phi^{\al,l}\Phi^{\bt,m}{\bf \bar{5}}_H\.\nn\\&&~~~~~+\left.y^e_{\al\bt}\epsilon_{il}\epsilon_{jm}{\bf 10}^{A,i}{\bf \bar{5}}^{B,j}\Phi^{\al,l}\Phi^{\bt,m}{\bf \bar{5}}_H \],
\eeqa
with the superscript $'i'$ denoting $SU(2)_H$ doublets index and the superscript $'s'$ denoting $SU(2)_H$ singlet (from the third generation).
The VEVs of two $SU(2)_F$ doublet higgs fields are aligned to lie on the upper and lower component of the corresponding doublet, respectively.
After $\Phi^{\al,i}$ acquires VEVs, the mixing between the third generation with the first two can be generated. Unlike some flavor $U(2)_H$ models\cite{su2-simple}, this general form of yukawa couplings will no longer be symmetric and all $(i,3)$ or $(3,i)$ entries [with $(i=1,2)$] of the yukawa matrices are non-vanishing. Thus it will not cause unpleasant mass and CKM-mixing relations.

The expansion of zero modes can then be written as
\beqa
\(\bea{c}\la^a_1(++)\\\la^a_2(--)\eea\)(x^\mu,y)=\sum\limits_{n}e^{-i\al \sigma_2 y /R}\(\bea{c}\la_{1n}^a\cos\[\f{n y}{R}\]\\\la^a_{2n}\sin(n+1)\f{y}{R}\eea\),
\eeqa
for gauginos.  Similarly for the Higgs sector we have
\beqa
\left(\bea{ccc} \phi_1~&\phi_2~&\phi_3  \\
\phi_1^{c\da}&\phi_2^{c\da}&\phi_3^{c\da}\eea \right)(x^\mu,y)&=&e^{- i\f{\al}{R}\sigma_2 y}\left(\bea{ccc} \phi_1[+,\pm]~&\phi_2[+,\mp]~&\phi_3[-,\mp] \\
\phi_1^{c\da}[-,\mp]&\phi_2^{c\da}[-,\pm]&\phi_3^{c\da}[+,\pm]\eea \right)(x^\mu,y)e^{i 2\f{\gamma y}{R} T^2} ,\nn\,\\
\left(\bea{ccc} \tl{\phi}_1~&\tl{\phi}_2~&\tl{\phi}_3  \\
\tl{\phi}_1^{c\da}&\tl{\phi}_2^{c\da}&\tl{\phi}_3^{c\da}\eea \right)(x^\mu,y)&=&\left(\bea{ccc}  \tl{\phi}_1[+,\pm]~&\tl{\phi}_2[+,\mp] ~&\tl{\phi}_3[-,\mp]\\
\tl{\phi}_1^{c\da}[-,\mp]&\tl{\phi}_2^{c\da}[-,\pm]&\tl{\phi}_3^{c\da}[+,\pm]\eea \right)(x^\mu,y)e^{i 2\f{\gamma y}{R} T^2},
\eeqa
with the decomposition depending on the orbifold projection
\beqa
\psi[+,+](x^\mu,y)&=&\sum\limits_{n=0}^\infty\f{1}{\sqrt{2^{\delta_{n,0}}\pi R}}\psi_{+,+}^n(x^\mu)\cos n\f{y}{R},\nn\\
\psi[+,-](x^\mu,y)&=&\sum\limits_{n=0}^\infty\f{1}{\sqrt{\pi R}}\psi_{+,-}^n(x^\mu)\cos (n+\f{1}{2})\f{y}{R},\nn\\
\psi[-,+](x^\mu,y)&=&\sum\limits_{n=0}^\infty\f{1}{\sqrt{\pi R}}\psi_{-,+}^n(x^\mu)\sin(n+\f{1}{2})\f{y}{R},\nn\\
\psi[-,-](x^\mu,y)&=&\sum\limits_{n=0}^\infty\f{1}{\sqrt{\pi R}}\psi^n_{-,-}(x^\mu)\sin(n+1)\f{y}{R} .
\eeqa

 After we integrate out the heavy Kaluza-Klein modes, we can obtain the contributions of
the SUSY soft parameters from the twisting boundary conditions
\beqa
\Delta{\cal L}&\supseteq& -\f{1}{2}\f{\al}{R}\( \tl{G}^a\tl{G}^a+\tl{W}^a\tl{W}^a+\tl{B}\tl{B}\)-\(\f{\al^2}{R^2}+4\f{\gamma^2}{R^2}\)\({h}_u^2+{h}_d^2\)\nn\\
&+& \f{4\al \gamma}{R^2} {h}_u {h}_d - \f{2\gamma}{R} \tl{h}_u\tl{h}_d - \(\f{\al^2}{R^2}+\f{\beta^2}{R^2}+\f{\gamma^2}{R^2}\)\sum\limits_{i=1}^2\(|\tl{Q}^i_L|^2+|\tl{U}^i_R|^2+|\tl{D}^i_R|^2\)\nn\\
&-& \(\f{\al^2}{R^2}+\f{\beta^2}{R^2}+\f{\gamma^2}{R^2}\)\sum\limits_{i=1}^2\(|\tl{L}^i_L|^2+|\tl{E}^i_R|^2\),
\eeqa
with $i=1,2$ being the family index.
In order to have the soft parameters at TeV scale,
we require $\al,\gamma\ll1$ , $\al/R \sim {\cal O}({\rm TeV})$
and ${\gamma}/{R}\sim {\cal O}~(100{\rm ~GeV})$.
An $SU(3)_c$ triplet Higgs field survives the orbifold projection.
Flavor symmetry guarantees that this triplet is inert.
For simply, we can add heavy brane mass terms for the $SU(3)_c$ triplet and
integrate out this field so that they do not appear in the low energy spectrum.
In fact, there is an alternative possibility concerning the boundary condition
of $\phi_2$. We can choose the boundary conditions for $\phi_2$ so that another
$SU(2)_L$ doublet will survive orbifold projection. We will leave this possibility
in our subsequent works.

\section{(Radiative) Natural SUSY with a large $A_t$ term}
\label{sect-2}
In natural SUSY the first/second generation squarks and sleptons are very heavy so that they
are beyond the LHC reach and also possibly heavy enough to provide a (partial) decoupling
solution to the SUSY flavor and CP problems. Naturalness requires that the third generation
sfermions are not too heavy. In order to have light third generations in our scenario,
we can put the third generation matter contents in the $y=0$ brane.
Thus there are no boundary breaking contributions to the sfermions of the third generation.
The third generation squark masses can be generated by gaugino loops:
\beqa
m_{\tl{f}}^2=\sum\limits_{i}\f{g_i^2}{16\pi^2}M_{\la_i}^2=\sum\limits_{I}\f{g_U^2}{16\pi^2} (M_\la^U)^2,
\eeqa
where $'I'$ denotes the number of interaction types involved in the loops.
This predicts that the stop masses are heavier than other sfermions of the third generation.
At the same time, the contributions of gaugino loops to the first two generations are subleading.
The soft masses for the third generation at the compactification scale are
\beqa
\label{sfermionthree}
m^2_{\tl{Q}_L^3}&=&\f{121}{60}\f{g_U^2}{16\pi^2}\f{\al^2}{R^2}~,~~m^2_{\tl{t}_L^c}=\f{19}{15}\f{g_U^2}{16\pi^2}\f{\al^2}{R^2}~,~~m^2_{\tl{b}_L^c}=\f{16}{15}\f{g_U^2}{16\pi^2}\f{\al^2}{R^2}~,~\nn\\
m^2_{\tl{L}_L^3}&=& \f{23}{20}\f{g_U^2}{16\pi^2}\f{\al^2}{R^2}~,~~~~m^2_{\tl{\tau}_L^c}=\f{3}{5}\f{g_U^2}{16\pi^2}\f{\al^2}{R^2}~,~~m^2_{\tl{\nu_\tau}_L^c}=0~.~
\eeqa
We can see that in our scenario the sfermion mass matrices are flavor blind and proportional to the identity matrix in the family space. We will see soon that the trilinear scalar couplings are also proportional to the relevant yukawa couplings.

While the advantages of natural SUSY are obvious (low EWFT, decoupling solution to SUSY flavor
and CP problems), some apparent problems seem to arise.
First, the sub-TeV top squarks usually lead to $m_h$
in the 115-120 GeV range, below 125 GeV.
The approximate one-loop formula for the Higgs mass \cite{Carena} is given by
\beqa
\label{higgs}
m_{h}^2=m_Z^2\cos^22\beta+\f{3m_t^4}{4\pi^2v^2}\[\ln\(\f{M_S^2}{m_t^2}\)+\f{X_t^2}{M_S^2}\(1-\f{X_t^2}{12M_S^2}\)\],
\eeqa
with
\beqa
X_t\equiv A_t-\mu\cot\beta,~~~~~M_S^2=m_{\tl{t}_1}m_{\tl{t}_2}.
\eeqa
An interesting observation is that the Higgs mass, as a function of $X_t/M_S$,
is maximal with the maximal mixing scenario $X_t/M_S=\sqrt{6}$ \cite{Wymant}.
So a large trilinear term $A_t$ can help to push up the Higgs mass.

In our scenario the trilinear terms can be generated by SUSY breaking boundary conditions.
The Lagrangian on the $y=0$ brane gives
\beqa
{\cal L}\supseteq \int d^2\theta \delta(y)\[y_2F_3^\pr \bar{f}_3^\pr \phi_3^c+ y_3 F_3^\pr F_3^\pr \phi_1\],
\eeqa
which leads to
\beqa
A_t=-y_3\f{\al}{R}\sim -\f{\al}{R}
\eeqa
after the F-term of the bulk hypermultiplet fields $\phi_1$,
which is $F_{\phi_1}=-(\pa_{\phi_1}W)^*-\pa_y \phi_1^c$, is substituted into
the superpotential. We should note that the trilinear term $A_t$ is independent
of $\gamma$ even when such an additional flavor symmetry is present.
At the compactification scale, the ratio
\beqa
\f{X_t}{M_S}\approx \f{A_t}{\sqrt{m_{\tl{t}_1}m_{\tl{t}_2}}}\sim -4\pi.
\eeqa
is fairly large. The absolute value of $A_t$ tends to increase with Renormalization Group Equation
running to lower scale which would give an even larger $X_t/M_S$.
So our scenario naturally gives a large $A_t$ to explain the observed 125 GeV Higgs.
On the other hand, the maximal mixing requires a not too large $A_t$ value.
So we will give a moderately large $A_t$ by introducing non-minimal Kahler potential.

We will introduce non-renormalizable K\"ahler terms for the two Higgs fields ($\Phi_1,\Phi_2$) in
${\bf 5}$ representations of SU(5). We can introduce a more general K\"ahler potential form
\beqa
K=\phi^\da e^{2gV}\phi+b_S\f{S+S^\da}{2M_*}\phi^\da e^{2gV}\phi~, \eeqa
with $S$ being a gauge singlet chiral field.
After the singlet $S$ develops a vacuum expectation value (vev), we get
\beqa \label{generalkahler}
K&{=}&(1+b_S\f{<S>}{M_*})\phi^\da e^{2gV}\phi.
 \eeqa
Because $S$ is a gauge singlet, it can acquire a vev of order $M_*$.

After orbifold projection and integrating out the heavy modes, we can obtain the wave function
normalization for $h_u$ and $h_d$
\beqa
Z_{h_u}&=&1+b_S^u\f{<S>}{M_*},~~
Z_{h_d}=1+b_S^d\f{<S>}{M_*}. \eeqa
Thus the mass terms appeared in previous section contributed from boundary conditions are
rescaled as
\beqa
\label{free1}
m_{h_u}^2&=&\f{1}{Z_{h_u}R^2}(\al^2+4\gamma^2)~,\nn\\
m_{h_d}^2&=&\f{1}{Z_{h_d}R^2}(\al^2+4\gamma^2)~,
\eeqa
as well as the terms for $\mu$ and $B\mu$
\beqa\label{free2}
\mu=\f{2\gamma}{\sqrt{Z_{h_u}Z_{h_d}}R},\nn\\
B\mu=\f{-4\al\gamma}{\sqrt{Z_{h_u}Z_{h_d}}R^2}.
\eeqa
We adopt the choice that $m_{h_u}^2 < m_{h_d}^2$ with $Z_{h_u}>Z_{h_d}$.
Such rescaling changes the UV input to $m_{h_u}^2\neq m_{h_d}^2$ to avoid possible problems
related to radiative electroweak gauge symmetry breaking which appears in mSUGRA
and GMSB.

With the previous non-minimal kinetic mass terms for $h_u$\footnote{There are also possible
brane localized kinetic terms for the third generation. Normalizing the kinetic term can also
contribute an additional factor to the trilinear coupling.}, the trilinear coupling can be
rescaled as
\beqa
A_t^\pr=\f{A_t}{\sqrt{Z_{h_u}}}=-y_t\f{\al}{\sqrt{Z_{h_u}} R}.
\eeqa
Thus the trilinear coupling could be rescaled to a moderately large value in our scenario.
Same arguments give $A_b^\pr=-y_b\f{\al}{\sqrt{Z_{h_d}}R}$.

Constraints from the LHC Higgs mass measurement suggest that the lower bound on $\tan\beta$
is $\tan\beta\gtrsim 3.5$ \cite{maximalmixing}. For such a large $\tan\beta$, successful
electroweak symmetry breaking gives the tree-level relation
\beqa
\label{ewsb2}
\f{m_Z^2}{2}=\f{\tan^2\beta m_{h_u}^2-m^2_{h_d}}{1-\tan^2\beta}-|\mu|^2\approx m^2_{h_u}-|\mu|^2.
\eeqa
Naturalness condition requires $m^2_{h_u}(M_{SUSY})\sim \mu^2(M_{SUSY})\sim M_Z^2/2$.
This requirement can be estimated to be
\beqa
\f{\al}{\ga}\approx\f{Z_{h_u}}{Z_{h_d}} ,
\eeqa
which can be relaxed if loop corrections are taken into account.
The electroweak symmetry breaking also requires
\beqa
|B\mu|^2>(|\mu|^2+m_{h_u}^2)(|\mu|^2+m_{h_d}^2).
\eeqa
In our scenario, it can be written as
\beqa
\f{4\al^2\ga^2}{{Z_{h_u}}{Z_{h_d}}R^4}>\f{\al^2(\al^2+4\gamma^2)}{Z_{h_u}^2 Z_{h_d}^2 R^4},
\eeqa
which can be simplified as
\beqa
 4 Z_{h_u} Z_{h_d}> \f{\al^2+4\gamma^2 }{\ga^2}.
\eeqa

In summary, our scenario has the following free parameters:
\bit
\item The parameters $\al$, $\bt$ (with $\bt\gg\al$ to evade flavor constraints) and $\ga$
related to SUSY breaking boundary conditions so that $\al/R\sim {\cal O}({\rm TeV})$.
Besides, natural SUSY requires a light Higgsino, which amounts to
$\gamma/R\sim {\cal O} ({\rm TeV})$ value after rescaling.
\item The Higgs field wavefunction normalization $Z_{h_u}$ and $Z_{h_d}$.
Naturalness condition requires $Z_{h_u}> Z_{h_d}>1$.
\eit
Then the SUSY parameters at the GUT scale are related to the above free parameters as
\bit
\item  The gaugino masses: $m_{1/2}=\f{\al}{R}$.
\item  The sfermion masses for the first two generations given by $m_0^2=\f{1}{R^2}(\al^2+\bt^2)$.
\item  The Higgs soft mass terms: $m_{h_u}^2=\f{\al^2+4\gamma^2}{Z_{h_u}R^2}$,
        $m_{h_d}^2=\f{\al^2+4\gamma^2}{Z_{h_d}R^2}$.
\item  The $\mu-B\mu$ term: $\mu=\f{2\ga}{\sqrt{Z_{h_u}Z_{h_d}}R}$,
          $B\mu=\f{-4\al\ga}{\sqrt{Z_{h_u}Z_{h_d}}R^2}$.
\item  The trilinear soft terms for the third generation:
       $A_t=-y_t\f{\al}{\sqrt{Z_{h_u}}R}$, $A_b=-y_b\f{\al}{\sqrt{Z_{h_d}}R}$.
\item  The sfermion masses for the third generation shown in Eq.(\ref{sfermionthree}).
\eit

\section{Viable parameter space}
The above soft supersymmetric parameters are obtained at the compactification scale.
Low energy soft SUSY spectrum can be obtained by solving the Renormalization Group
Equation at the weak scale. This procedure is done with the code {\bf SuSpect 2.4.1} \cite{suspect}.
The unification of the matter contents eliminates many free parameters.
With the remained free parameters, this scenario is greatly constrained by various experiments.
In our study we consider the following experimental constraints on the parameter space:
\begin{itemize}
\item[(1)] The $b \rightarrow s\gamma$ decay branching ratio
BF($b \rightarrow s\gamma$) = $(3.55 \pm 0.26)\times 10^{-4}$ \cite{bsgamma}.
We require the theoretical value to be in the $3\sigma$ range of the experimental data.
\item[(2)] The $B_s \rightarrow \mu^{+}\mu^{-}$ from LHCb measurement \cite{lhcb}:
BF($B_s \rightarrow \mu^{+}\mu^{-}$) = $3.2^{+1.5}_{-1.2}\times 10^{-9}$.
We require the theoretical value to be in the $3\sigma$ range.
\item[(3)]  The dark matter relic density $\Omega h^2=0.1126 \pm 0.0036$
from WMAP \cite{wmap} and $\Omega h^2=0.1199 \pm 0.0027$ from Planck \cite{planck}.
We require the relic density of the neutralino dark matter satisfy the $2\sigma$ upper bound.
\item[(4)] The XENON100(2012) constraints on the dark matter scattering off the nucleon \cite{xeon}.
\end{itemize}
Note that we do not impose $(g-2)_{\mu}$ constraints \cite{gmuon-2} in our scenario.
Due to heavy $\tilde{\mu}_{1,2}$ and $\tilde{\nu_{\mu}}$, the SUSY contribution
to $(g-2)_{\mu}$ is small in our scenario.

Under above constraints, we scan over the parameter space in our scenario with the
code {\bf MicrOmega 1.3} \cite{micromega} and {\bf SUSY\_FLAVOR v2.02} \cite{Crivellin} .
In general, the parameters related to the electroweak symmetry breaking sector contain
$m_{h_u}^2$, $m_{h_d}^2$, $|\mu|^2$, $B_{\mu}$ and $\tan\beta$.
Successful radiative electroweak symmetry breaking requires these parameters to
satisfy  (\ref{ewsb2}) and
\beqa
\label{ewsb1}
\sin2\beta=-\f{2 B\mu}{m_{h_\mu}^2+m_{h_d}^2+2|\mu|^2}.
\eeqa
Thus there remained essentially three parameters, which are taken as
$\tan\beta$, $\mu$ and $m_A$ at weak-scale.
These parameters are related to $m_{h_u}^2$, $m_{h_d}^2$, $|\mu|^2$ and $B_{\mu}$ by
\beqa
m_A^2&=&\f{2B_{\mu}}{\sin2\beta}=2|\mu|^2+m_{h_u}^2+m_{h_d}^2~.
\eeqa
Also, these three free parameters are related to $Z_{h_u}$, $Z_{h_d}$ and $\gamma$
by Eqs.(\ref{free1}) and (\ref{free2}).
In natural SUSY, the soft masses of the first two generations are of order
${\cal O}(10){\rm ~TeV}$. So we can choose $\beta/R=10$ {\rm ~TeV}.
We search for natural SUSY solutions in the parameter space in our scenario
and perform a random scan in the following ranges:
\begin{eqnarray}
&&          100{\rm ~GeV} <  m_{1/2} < 4 {\rm ~TeV}, \nonumber \\
&&          100{\rm ~GeV} <  -\mu   < 150{\rm ~GeV},\nonumber \\
&&           150{\rm ~GeV} <  m_A   < 1.5{\rm ~TeV}, \nonumber \\
&&           3< \tan\beta < 50, \nonumber \\
&&           -4 < A_{t}/m_{1/2} < 4.
\end{eqnarray}

In our scan, we choose the top quark mass to be 175 GeV. The $SU(2)_L$ and $U(1)_Y$ gauge
couplings at the weak scale lead to the following GUT-scale boundary conditions
\begin{eqnarray}
\alpha_{GUT} = \f{1}{24.3},~ ~M_{GUT} = 2.0\times 10^{16}{\rm~GeV} , ~~
y_t(M_{GUT}) = 0.51,~~ y_b(M_{GUT}) = 0.054.
\end{eqnarray}
The unification scale is determined by $g_2(M_{GUT}) = g_1(M_{GUT}) = g_{GUT}$.
Note that due to the relatively large uncertainty of
$SU(3)_c$ gauge coupling at the electroweak scale,
the value of $\al_s$ at the electroweak scale is usually regarded as a GUT prediction
by the RGE running of $\al_{GUT}$.

Natural SUSY requires light stops.
 So we require $m_{\tilde{t}_1}<1.5$ TeV and
$m_{\tilde{t}_2}< 2$ TeV. In our previous discussion, the stop masses are generated
via gaugino loops and satisfy the relation
$M^2_{\tl{Q}_L^3}\approx 2\f{g_U^2}{16\pi^2}\f{\al^2}{R^2}$
and $M^2_{\tl{t}_L^c}= \f{19}{15}\f{g_U^2}{16\pi^2}\f{\al^2}{R^2}$.
Thus, heavy gauginos will in general lead to heavy stop masses.
So from stop masses we anticipate an upper limit for the universal $m_{1/2}$.
 In radiative natural supersymmetry, the condition of light stop can be relaxed without spoiling the naturalness conditions.
 This is welcome since it is not easy for natural supersymmetry to give 125 GeV higgs even in case of maximal mixing.

\begin{figure}[htb]
\begin{center}
\includegraphics[width=13cm]{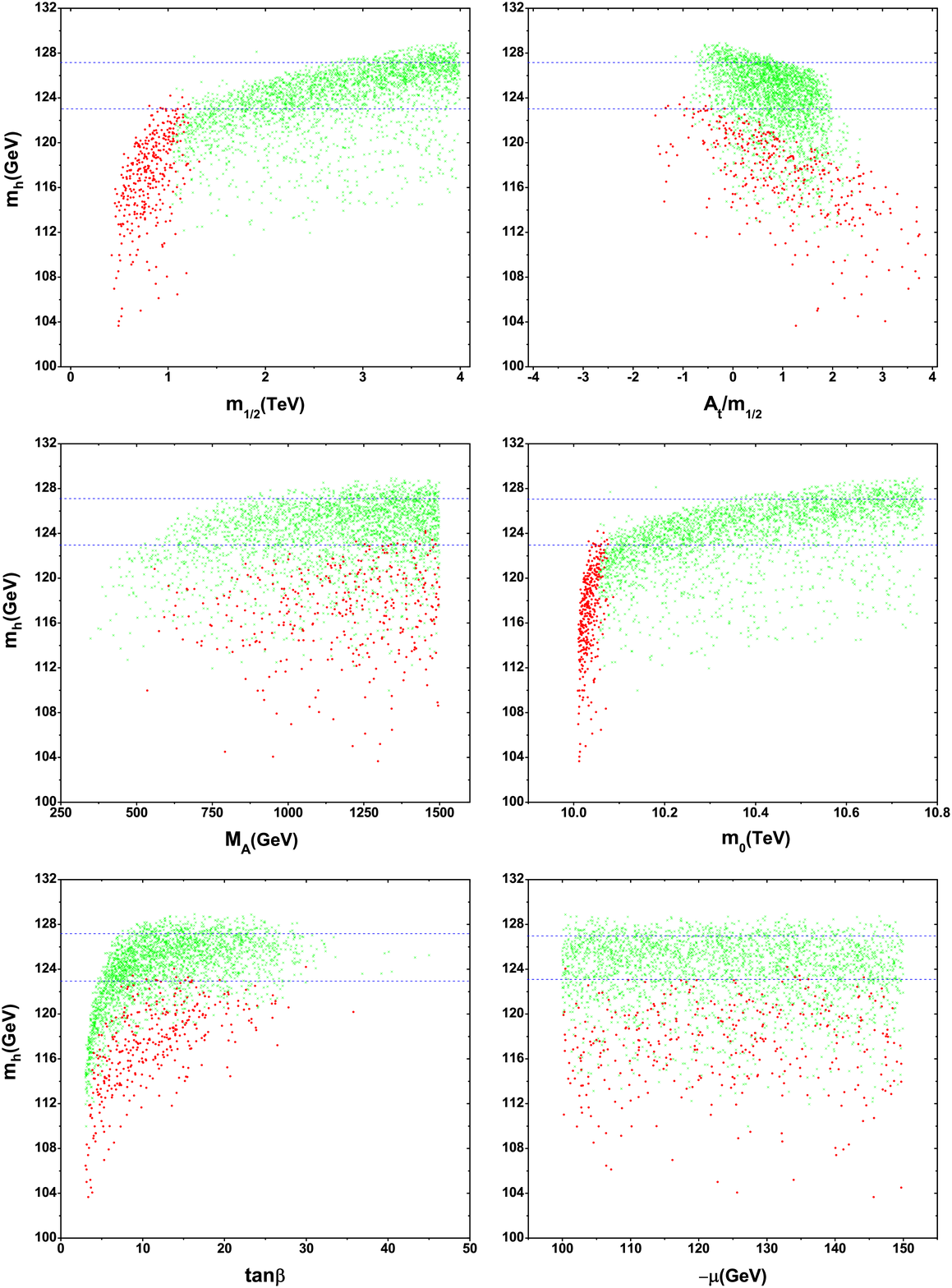}
\end{center}
\vspace{-0.7cm} \caption{Scatter plots of the survived samples in the parameter space.
The bullets (red) satisfy all considered constraints including
$m_{\tilde{t}_1}<1.5$ TeV and $m_{\tilde{t}_2}< 2$ TeV.
The crosses (green) satisfy the constraints (1-4) but do not satisfy
$m_{\tilde{t}_1}<1.5$ TeV or $m_{\tilde{t}_2}< 2$ TeV. The two horizontal lines
show the Higgs mass range of 123-127 GeV.}
 \label{fig1}
\end{figure}

\begin{figure}[htb]
\begin{center}
\includegraphics[width=13cm]{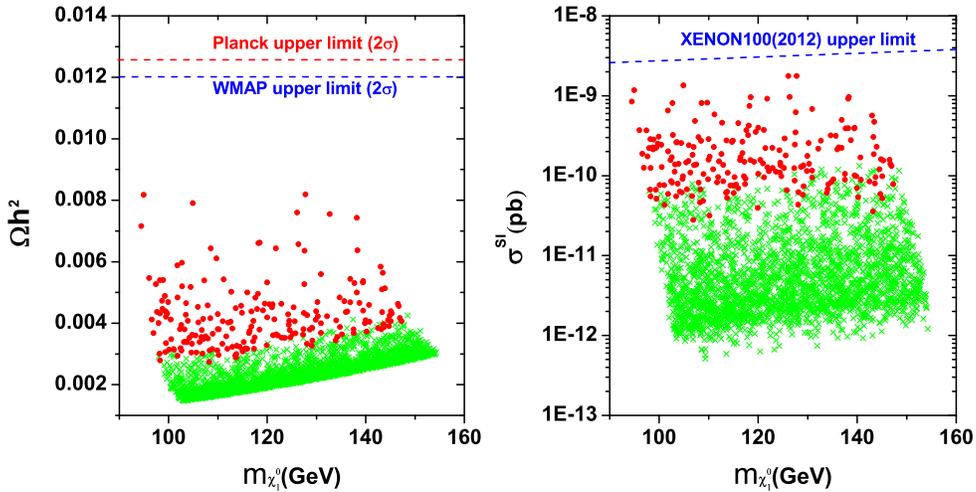}
\end{center}
\vspace{-0.7cm}
\caption{Same as Fig.1, but showing the neutralino relic density
and the spin-independent neutralino-proton scattering cross section
versus the neutralino mass.  The horizontal lines show the
$2\sigma$ upper limits from WMAP, Planck and XENON100(2012).}
\label{fig2}
\end{figure}

\begin{figure}[htb]
\begin{center}
\includegraphics[width=10cm]{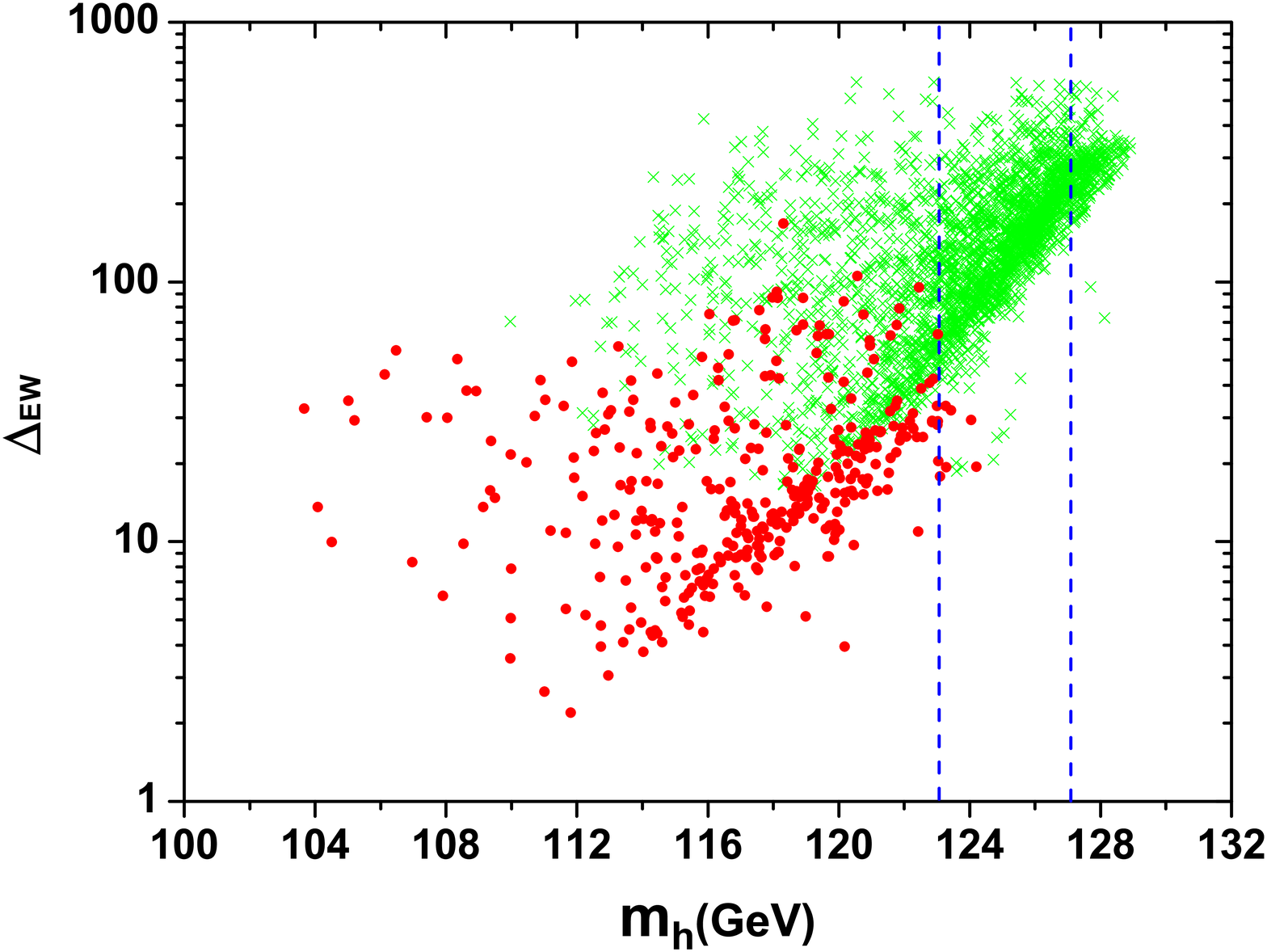}
\end{center}
\vspace{-0.7cm}
\caption{Same as Fig.1, but showing the fine-tuning extent versus the SM-like Higgs mass.
 The two vertical lines show the Higgs mass range of 123-127 GeV.}
 \label{fig3}
\end{figure}

In Fig.1 we present the scatter plots for the survived samples.
The samples which satisfy the constraints (1-4) but do not satisfy
$m_{\tilde{t}_1}<1.5$ TeV or $m_{\tilde{t}_2}< 2$ TeV are also displayed.
We can see that the allowed $m_{1/2}$ value is below 1.5 TeV, which
is mainly from the requirement of light stops.
On the other hand, since low stop masses are difficult to push up the higgs boson mass,
only a few survived samples can give a SM-like higgs boson in the range of 123-127 GeV.
However, if we do not impose the stop mass upper limits,
a 125 GeV SM-like higgs can be obtained easily \cite{Arganda}, as shown by this figure.
As we noted before, relatively heavier stop mass is natural in radiative natural supersymmetry scenario.
This figure also shows the allowed ranges for other parameters.
We can see that a negative $A_t$ can not be too large.
Besides, a large $M_A$ and a moderate $\tan\beta$ is favored due to the
flavor physics constraints.

We show in Fig.2 the neutralino relic density
and the spin-independent neutralino-proton scattering cross section
versus the neutralino mass.
Due to the low value of $\mu$, the neutralino $\chi^0_1$ is higgsino-like
which has large annihilation cross section and thus low relic density
In natural SUSY the neutralino dark matter is higgsino-like
and its relic density is not sufficient to explain the measured value.
This means that dark matter cannot be solely made of the neutralino
and other components should exist.

We also see from fig.2 that the samples with light stops
give a larger relic density than the samples with heavy stops.
The reason is that a small stop mass means a smaller $m_{1/2}$ which further
induces a light $\tilde{\tau}$. A light $\tilde{\tau}$ can lower the
neutralino annihilation cross section through $t$-channel and thus
raise the neutralino relic density.

 There are various approaches to introduce an additional dark matter candidate besides the lightest neutralino.
For example, we can introduce Peccei-Quinn mechanism to solve the strong CP problem. The axion with respect to PQ symmetry breaking will be dark matter candidates.
In supersymmetric cases, axion will be promoted into a supermultiplets involving saxion and axino.
Dark matter could consist of two particles: an axion-higgsino admixture.
Axino decay can alter the relic abundance of neutralino dark matter which depends on the lifetime and the energy density of axino.
The neutralinos are thermally produced, they can also be produced by thermal production followed by cascade decays of axino at high reheating temperature. The late decay of axinos into
higgsinos could cause a re-annihilation of neutralinos at a temperature below freeze-out, substantially increasing the relic abundance.
Because of large higgsino annihilation cross section in our scenario, heavy axino decay can lead to reasonable dark matter density of the lightest neutralino and enough radiation to dilute the
gravitinos\cite{axino}.

At last, we show the fine-tuning extent versus the Higgs mass in Fig.3.
Going beyond the tree-level expression (\ref{ewsb2}), the loop-level
minimization condition of the Higgs potential is
\begin{eqnarray}\label{loopewsb}
\frac{M^2_{Z}}{2}=\frac{(m^2_{H_d}+\Sigma_{d})-(m^2_{H_u}+
\Sigma_{u})\tan^{2}\beta}{\tan^{2}\beta-1}-|\mu|^{2},
\end{eqnarray}
where $\Sigma_{u}$ and $\Sigma_{d}$ are the radiative corrections to the Higgs potential and the
dominant contribution to the $\Sigma_{u}$ is given by
\begin{eqnarray}
\Sigma_u \sim \frac{3Y_t^2}{16\pi^2}\times m^{2}_{\tilde{t}_i}
\left( \log\frac{m^{2}_{\tilde{t}_i}}{Q^2}-1\right)\;.
\label{rad-corr}
\end{eqnarray}
The fine-tuning extent is defined by \cite{Baer}
\beqa
\Delta_{EW}=\f{\max|C_i|}{M_Z^2/2}~,
\eeqa
where $C_i (i=h_u,h_d,\mu,\Sigma_u,\Sigma_d)$ denotes each term in the right side
of $(\ref{loopewsb})$.
From Fig.3 we see that in general light-stop points correspond to a low fine-tuning extent.
Note that there are some heavy-stop points which can give a low fine-tuning
extent (below 30). This means that in our scenario
the radiative natural SUSY can be realized.

\section{Conclusions}
\label{sec-6}
 In this paper, we proposed a realistic 5D orbifold GUT model that can reduce
to (radiative) natural supersymmetry in the low energy. Supersymmetry as well as
gauge symmetry are broken by the twist boundary conditions. We found that it is
non-trivial to introduce other flavor symmetry other than the $SU(2)_R$ R-symmetry.
The tension between the limited number of parameters and the successful electroweak
symmetry breaking can be  ameliorated by introducing non-minimal Kahler potentials.
A large trilinear term $A_t$, which is necessary to give a 125 GeV higgs is
naturally predicted in our scenario. A scan under current experimental constraints shows
that our model can realize natural (or radiative natural) supersymmetry. Only radiative natural supersymmetry can naturally lead to 125 GeV higgs.
Additional dark matter candidates other than neutralino (like axion) are needed to provide enough dark matter relic density.
Besides, relatively large stop masses are necessary to give realistic higg mass in most of the parameter spaces.

\begin{acknowledgments}
We are grateful to the referee for discussions. Fei Wang and Chengcheng Han acknowledge a joint-research
program between Institute of Theoretical Physics and Zhengzhou University.
This research was supported by the
National Natural Science Foundation of China under grant numbers 11105124,11105125,
11275245, 10821504 and 11135003, by the Project of Knowledge Innovation Program (PKIP) of Chinese Academy of Sciences under grant number KJCX2.YW.W10.
\end{acknowledgments}


\begin{thebibliography}{99} \vspace{-1mm}
\bibitem{atlas} G. Aad et al.(ATLAS Collaboration), Phys Lett.B 710(2012) 49.

\bibitem{cms} S. Chatrachyan et al.(CMS Collaboration), Phys Lett.B 710(2012) 26.

\bibitem{Ellis:1990zq}
J.~R.~Ellis, S.~Kelley and D.~V.~Nanopoulos,
Phys.\ Lett.\ B {\bf 249}, 441 (1990);
Phys.\ Lett.\ B {\bf 260}, 131 (1991);
U.~Amaldi, W.~de Boer and H.~Furstenau,
coupling
Phys.\ Lett.\ B {\bf 260}, 447 (1991);
P.~Langacker and M.~X.~Luo,
$\rho_{0}$,
Phys.\ Rev.\ D {\bf 44}, 817 (1991).
\bibitem{Georgi:1974sy}
H.~Georgi and S.~L.~Glashow,
Phys.\ Rev.\ Lett.\ {\bf 32}, 438 (1974).

\bibitem{so10} H. Georgi, ``Particles And Fields: Williamsburg 1974.
AIP Conference Proceedings No. 23'', Editor C.~E.~Carlson;
H.~Fritzsch and P.~Minkowski,
Annals Phys.\ {\bf 93}, 193 (1975);
H.~Georgi and D.~V.~Nanopoulos,
Nucl.\ Phys.\ B {\bf 155}, 52 (1979).


\bibitem{kawamura} Y. Kawamura, Prog. Theo. Phys 103 (2000) 613;
                    hep-ph/0012125; hep-ph/0012352.

\bibitem{at} G. Altarelli and F. Feruglio, Phys. Lett. B511, 257 (2001);

\bibitem{hall} L. Hall and Y. Nomura, Phys. Rev. D64 (2001) 055003.

\bibitem{hebecker:2001wq} A. Hebecker, J. March-Russell, Nucl. Phys. B625, 128 (2002).

\bibitem{hebecker:2001jb} A. Hebecker, J. March-Russell, Nucl. Phys. B613, 3 (2001).

\bibitem{Li:2001qs} T.~Li,  Phys.\ Lett.\  B {\bf 520}, 377 (2001).

\bibitem{Li:2001wz} T.~Li, Nucl.\ Phys.\  B {\bf 619}, 75 (2001).

\bibitem{fei2} C. Balazs, T. Li, F. Wang, J. M. Yang, JHEP 0909,015(2009).

\bibitem{fei3} C. Balazs  T. Li, Dimitri V. Nanopoulos, F. Wang,  JHEP02(2010)096.

\bibitem{li2010} T.Li, Dimitri V. Nanopoulos, JHEP10(2011)090.

\bibitem{adscft} J. Maldacena, Adv. Theor. Math. Phys. 2, 231(1998).

\bibitem{rs} L. Randall, R. Sundrum, Phys. Rev. Lett. 83, 3370 (1999).

\bibitem{naturalsusy}
R. Kitano and Y. Nomura, Phys. Lett. B631, 58 (2005); Phys. Rev. D73, 095004  (2006);
H.~Baer, V.~Barger, P.~Huang and X.~Tata, JHEP {\bf 1205} (2012) 109;
J.~Cao {\it et al.}, JHEP {\bf 1211} (2012) 039.

\bibitem{cao}  J.~Cao {\it et al.}, JHEP {\bf 1203}, 086 (2012); JHEP {\bf 1210}, 079 (2012);
                                    Phys. Lett. B {\bf 703}, 462 (2011).

\bibitem{rnaturalsusy} H. Baer {\it et al.}, Phys. Rev. Lett. 109, 161802 (2012).

\bibitem{mSUGRA}
A.~H.~Chamseddine, R.~L.~Arnowitt and P.~Nath,
Phys.\ Rev.\ Lett.\ {\bf 49}, 970 (1982);
H.~P.~Nilles,
Phys.\ Lett.\ B {\bf 115}, 193 (1982);
L.~E.~Ibanez,
Phys.\ Lett.\ B {\bf 118}, 73 (1982);
R.~Barbieri, S.~Ferrara and C.~A.~Savoy,
Phys.\ Lett.\ B {\bf 119}, 343 (1982);
H.~P.~Nilles, M.~Srednicki and D.~Wyler,
Phys.\ Lett.\ B {\bf 120}, 346 (1983);
J.~R.~Ellis, D.~V.~Nanopoulos and K.~Tamvakis,
Phys.\ Lett.\ B {\bf 121}, 123 (1983);
J.~R.~Ellis, J.~S.~Hagelin, D.~V.~Nanopoulos and K.~Tamvakis,
Supergravity,'' Phys.\ Lett.\ B {\bf 125}, 275 (1983);
L.~J.~Hall, J.~D.~Lykken and S.~Weinberg,
Phys.\ Rev.\ D {\bf 27}, 2359 (1983).

\bibitem{fei4} C. Balazs, T. Li, D. V. Nanopoulos, F. Wang, JHEP09, 003 (2010).

\bibitem{fei5} F. Wang, Nucl. Phys. B851, 104 (2011).

\bibitem{nima2} N. Arkani-Hamed, L. Hall, D. Smith, N. Weiner, Phys. Rev. D63, 056003 (2001);
                N. Arkani-Hamed, T.Gregoire, J.Wacker, JHEP 0203, 055 (2002).

\bibitem{quiros}  A. Pomarol, M. Quiros, Phys. Lett. B438, 255 (1998);
                  M. Quiros, hep-ph/0302189.

\bibitem{nomura} R. Barbieri, L. Hall, Y. Nomura, Phys. Rev. D66, 045025 (2002);\\
	             R. Barbieri, L. Hall, Y. Nomura, Nucl.Phys.B624:63-80,2002;\\
                 H. Murayama, Y. Nomura, S. Shirai, K. Tobioka,  Phys.\ Rev.\ D86,115014(2012).

\bibitem{scherkschwarz} J. Scherk, J. H. Schwarz, Phys. Lett. B82, 60 (1979);
                        J. Scherk, J. H. Schwarz, Nucl. Phys. B153, 61 (1979);
                        E. Cremmer, J. Scherk, J. H. Schwarz, Phys. Lett. B84, 83 (1979).

\bibitem{susyboundary} I. Antoniadis, Phys. Lett. B246, 377 (1990);
                       I. Antoniadis, S. Dimopoulos, A. Pomarol, M. Qurios, Nucl. Phys. B544, 503 (1999);
                       R. Barbieri, J. Hall, Y. Nomura, Phys. Rev. D66, 045025 (2002).
\bibitem{pomarol} D. Marti,, Phys. \ ReV. \ D65, 105025(2001).

\bibitem{gersdorff} G. V. Gersdorff, A. Pomarol, Phys. \ ReV. \ D64, 064016(2002).

\bibitem{GW-mechanism} Walter D. Goldberger, Mark B. Wise, Phys. \ Rev. \ D60 (1999) 107505.
\bibitem{su2-global}M. Dine, A. Kagan, R. G. Leigh, Phys. \ Rev. \ D48:4269-4274,1993.

\bibitem{pomarol-ns} Emilian Dudas, Gero von Gersdorff, Stefan Pokorski, Robert Ziegler, [hep-ph/1308.1090].

\bibitem{su2-simple}R. Barbieri, G. R. Dvali and L. J. Hall, Phys. Lett. B377, 76 (1996).

\bibitem{Carena}
  M.~Carena, S.~Gori, N.~R.~Shah and C.~E.~M.~Wagner,
    JHEP {\bf 1203} (2012) 014.
\bibitem{Wymant}
  C.~Wymant,
  Phys.\ Rev.\ D {\bf 86} (2012) 115023.






\bibitem{maximalmixing} P. Draper, P. Meade, M. Reece, D. Shih, Phys. Rev. D85, 095007 (2012).

\bibitem{suspect} A. Djouadi, J. Kneur, G. Moultaka, Comput. Phys. Commun. 176, 426 (2007).

\bibitem{bsgamma} D. Asner {\it et al.} [Heavy Flavor Averaging Group], arXiv:1010.1589 [hep-ex].

\bibitem{lhcb} R. Aaij  {\it et al.} [LHCb Collaboration], Phys. Rev. Lett. 108, 231801 (2012).

\bibitem{wmap} J. Dunkley  {\it et al.} [WMAP Collaboration], Astrophys. J. Suppl. 180, 306 (2009).

\bibitem{planck} P. A. R. Ade et al. [Planck Collaboration], arXiv:1303.5076.

\bibitem{xeon} E. Aprile,  {\it et al.} [XENON100 Collaboration], arXiv:1207.5988 [astro-ph.CO].

\bibitem{gmuon-2}  M. Davier {\it et al.}, Eur. Phys. J. C 66, 1 (2010).
\bibitem{micromega} G. Belanger, {\it et al.}, Comput. Phys. Commun. 182, 842 (2011);
                    Comput. Phys. Commun. 174, 577 (2006);
                    Comput. Phys. Commun. 149, 103 (2002).

\bibitem{Crivellin}
  A.~Crivellin {\it et al.},  Comput.\ Phys.\ Commun.\  {\bf 184} (2013) 1004.

\bibitem{Arganda}
  E.~Arganda, J.~L.~Diaz-Cruz and A.~Szynkman,
  arXiv:1301.0708.


\bibitem{Baer}
  H.~Baer,
  arXiv:1210.7852 [hep-ph].
\bibitem{axino} Ki-Young Choi, Jihn E. Kim, Hyun Min Lee, Osamu Seto, Phys.Rev.D77:123501,2008
\end{thebibliography}
\end{document}